\documentclass[
reprint,
superscriptaddress,
nofootinbib,
amsmath,amssymb,
aps,
prd,
floatfix,
]{revtex4-1}

\usepackage{color}
\usepackage{graphicx}
\usepackage{dcolumn}
\usepackage{bm}
\usepackage{hyperref}

\hypersetup{
    colorlinks=true,
    linkcolor=blue,
    citecolor=blue,
    filecolor=black,
    urlcolor=blue,
}

\newcommand{\mpl}{M_\mathrm{Pl}}

\def \min {\mathrm{min}}

\def\vev#1{ \left\langle #1 \right \rangle }
\def\impc{\; \mathrm{Mpc}^{-1}}

\begin{document}

\title{Power spectrum oscillations from Planck-suppressed operators in effective field theory motivated monodromy inflation}

\author{Layne C. Price}
  \email{lpri691@aucklanduni.ac.nz}
  \affiliation{Department of Physics, University of Auckland, Private Bag 92019,  Auckland, New Zealand}

\date{\today}

\begin{abstract}

We consider a phenomenological model of inflation where the inflaton is the phase of a complex scalar field $\Phi$.  Planck-suppressed operators of $\mathcal O(f^5/M_\mathrm{pl})$ modify the geometry of the vev $\vev{\Phi}$ at first order in the decay constant $f$, which adds a first order periodic term to the definition of the canonically normalized inflaton $\phi$.
This correction to the inflaton induces a fixed number of extra oscillatory terms in the potential $V \sim \theta^p$.
We derive the same result in a toy scenario where the vacuum $\vev{\Phi}$ is an ellipse with an arbitrarily large eccentricity.
These extra oscillations change the form of the power spectrum as a function of scale $k$ and provide a possible mechanism for differentiating EFT-motivated inflation from models where the angular shift symmetry is a gauge symmetry.

\end{abstract}

\maketitle

\section{Introduction}

Single-field, slow-roll inflation models that generate detectable levels of primordial gravitational waves require a super-Planckian field displacement for the inflaton~\cite{Lyth:1996im}.
The constraints on the primordial power spectrum from \emph{Planck}~\cite{Ade:2013rta,Ade:2013ydc,Ade:2015lrj} remain broadly consistent with the predictions of large classes of these models, particularly those with $V \sim \phi^p$ for $0<p \lesssim 2$.
Knowledge of the detailed predictions of these models in the range $p \ll 2$ has become increasingly important as estimates of the tensor-to-scalar ratio $r$ have improved~\cite{Ade:2013zuv,Ade:2015tva,Ade:2015xua}.

However, Planck-scale effects should substantially alter the simple polynomial relationship $V \sim \phi^p$ over regions of field space that exceed $\Delta \phi \gtrsim 1$.
Monodromy is a mechanism that protects against these dangerous corrections by repeatedly wrapping the trajectory of the inflaton around a sub-Planckian region of a higher dimensional space, with an angular parameter $\theta$.
The mechanism relies on a corresponding change $\Delta V$ for each winding $\theta \to \theta + 2\pi$, with the global potential $V(\phi)$ defined by patching together branch cuts.  Large field excursions $\Delta \phi \gtrsim 1$ are then safe because the displacement in the ``natural'' parameter space remains small.

Importantly, monodromy inflation predicts detectable oscillatory features in the primordial power spectrum of curvature perturbations $P_\zeta(k)$~\cite{McAllister:2008hb,Flauger:2009ab}.  If the angular field $\theta$ is coupled to gauge fields, then non-perturbative instanton effects modify the potential $V \to V + g \sin (\theta_i)$.  In the absence of monodromy, this oscillatory contribution can drive natural inflation~\cite{Freese:1990rb}.  Combined with a monodromy effect in $V$, this contribution imprints oscillations in the power spectrum with a frequency that is related to the periodicity of the underlying theory through the axion decay constant $f$.  While the size of these oscillations is constrained by the WMAP~\cite{Peiris:2013opa} and \emph{Planck} data~\cite{Ade:2013rta,Easther:2013kla,Meerburg:2013cla,Meerburg:2013dla,Ade:2015lrj}, their detection would provide an unprecedented glimpse into the monodromy structure of the high energy theory.  Therefore, it is important to understand the exact nature of these oscillations and to identify any mechanisms that might change this prediction.

In this paper we study an alternative source of oscillations in the primordial spectra in monodromy inflation, where the angular degree of freedom $\theta$ is the phase of a complex scalar $\Phi$ with a vacuum expectation value (vev) of $\vev{\Phi} = \rho$.
Adding non-renormalizable terms to the action for $\Phi$ alters the shape of the vacuum so that $\rho \to \rho(\theta)$ and the effective decay constant obtains an angular dependence.
These higher order terms only contribute to the potential as $V_\mathrm{NR} \sim \mathcal O \left(f^5 \right)$, which is much smaller than unity for sub-Planckian decay constants $ f \ll 1$.  However, they change the form of the canonical kinetic energy term for $\Phi$ in the vacuum at first order in $f$.   Consequently, the canonically normalized inflaton $\phi$, which depends on the structure of the kinetic energy term for $\Phi$, obtains a quasi-periodic dependence at $\mathcal O(f^{m-4})$, where $m \ge 5$ is the energy dimension of the non-renormalizable operator.

Therefore, the leading order gravitational operators with $m=5$ add first order oscillatory corrections to any term in the scalar action that depends on the angular coordinate $\theta$.  If the shift symmetry in the vacuum $\theta \to \theta + \mathrm{const.}$ is broken, then the non-renormalizable operators add a set of sinusoidal functions to $V$ that have arbitrary relative phases.  These new terms are independent of the scale of any instanton effects derived from gauge couplings and result in a superposition of high frequency power spectrum oscillations.  These features must be taken into account when fitting the predictions of monodromy inflation to data.

Additionally, we consider a case where the structure of the vacuum $\vev{\phi}$ is shaped like an ellipse with an arbitrarily large eccentricity.
The coupling constant for the oscillatory terms are suppressed by a factor of at least $1/3$ compared to the angular symmetry breaking scale, even for a highly elliptical vev.
A flat prior probability on the eccentricity $P(\xi) \sim \mathcal U[0,1]$, which corresponds to a naturally large correction to the vev, gives sinusoidal terms in $P_\zeta(k)$ that are suppressed at the order of $\mathcal O(10^{-2})$ compared to the contribution from the chaotic potential $V \sim \theta^p$.  This gives power spectrum features that are still safely inside the current bounds from \emph{Planck} data~\cite{Easther:2013kla}.

Adding Planck-suppressed terms to the Lagrangian is motivated by an effective description of monodromy inflation at the energy scale $\rho \ll \mpl$.  However, we do require the existence of global symmetries in the low-energy theory, which is a non-trivial assumption that makes obtaining large-field inflation simple.  A more rigorous description of the low-energy theory might require significant fine-tuning in order to obtain a flat potential across super-Planckian field ranges.
Consequently, we assume that an angular symmetry breaking term is technically natural in a field theory that contains a global angular shift symmetry~\cite{Behbahani:2011it} and not include operators that would spoil the chaotic inflation behavior of the potential $V \sim \theta^p$.
We take a phenomenological approach and remain agnostic about a specific high energy scenario that realizes this symmetry.

Monodromy arising from a gauge symmetry is a ubiquitous feature of high energy theories.  In particular, string theory has many axions, which are angular parameters that are given a monodromy term by couplings to fluxes~\cite{Silverstein:2008sg,McAllister:2008hb,Kaloper:2008fb,Flauger:2009ab,Dong:2010in,Kaloper:2011jz,Kaloper:2014zba,McAllister:2014mpa}.  The monodromy potential  for string axions $V\sim \phi^p$ can have a variety of exponents, with $p=\{2/3,1,4/3,2\}$ giving predictions that match the \emph{Planck} posterior probabilities for the $\Lambda\mathrm{CDM}+r$ model within approximately $1$-$2 \sigma$~\cite{McAllister:2014mpa}.
While the primordial spectrum resulting from monodromy inflation with a broken global shift symmetry has an oscillatory contribution from each Planck-suppressed operator,
the simplest models with a gauged axion symmetry give a coherent, sinusoidal signal in $\log k$.  In principle, these two signals can be distinguished via cosmic microwave background or large-scale structure data.

Furthermore, although a large number of angular parameters $\theta_i$ could contribute to inflation in principle, we only consider the single-field case.  While many-field models can have simple predictions~\cite{Frazer:2013zoa,Easther:2013rva,Wenren:2014cga}, the variance of the predicted values of observables for monodromy inflation can be undefined for $p<3/4$~\cite{Price:2014ufa}.  Since small exponents $p$ are observationally favored, we will leave an analysis of many-field monodromy to future work.

This paper is organized as the following: Sect.~\ref{sect:vac} discusses how non-renormalizable operators alter the shape of the vacuum for a complex scalar with a Mexican hat type potential.  Sect.~\ref{sect:ellipse} calculates the canonically normalized inflaton for an arbitrarily elliptical vev.  Sect.~\ref{sect:pk} estimates the shape of the features in $P_\zeta(k)$ from this mechanism both analytically and numerically.  Sect.~\ref{sect:diff} compares the predicted spectra from Planck-corrected, EFT-motivated monodromy models to high energy axion monodromy scenarios.  Sect.~\ref{sect:concl} summarizes the findings.

\section{Modifying the geometry of the vacuum with irrelevant operators}
\label{sect:vac}

We start by considering a Lagrangian for a complex scalar field,
\begin{equation}
  \Phi = \rho \; e^{i \theta},
  \label{eqn:XXX}
\end{equation}
where the renormalizable terms in the action satisfy the canonical Mexican hat shape with
\begin{equation}
  \mathcal L = -\frac{1}{2} \partial_\nu \Phi^* \partial^\nu \Phi - \lambda \left( |\Phi|^2 -f^2 \right)^2.
  \label{eqn:mexhat}
\end{equation}
We assume that the dimensionless real parameter $\lambda$ is $\mathcal O(1)$ and that the decay constant is $f \ll 1$.
The vev $\vev{\Phi}=f$ is independent of the phase $\theta$ and is sub-Planckian for small decay constants.
In the vacuum, the radial component of $\Phi$ has the mass $m_\rho^2 = 4 \lambda f^2$.
If $\theta$ is coupled to gauge fields, then non-perturbative instanton effects induce a sinusoidal correction to Eq.~\eqref{eqn:mexhat}, which gives rise to natural inflation if $f \gg 1$~\cite{Freese:1990rb}.

However, monodromy inflation can be generated in the more natural $f \ll 1$ limit by
explicitly breaking the $\theta \to \theta + \mathrm{const}.$ global symmetry.  We do this by adding a term to Eq.~\eqref{eqn:mexhat} like
\begin{equation}
  V_\mathrm{ASB} = g \, \theta^p,
  \label{eqn:mono}
\end{equation}
where the subscript ASB denotes angular symmetry breaking.
We assume that the form of this term remains stable over regions $|\Phi| \ll 1$, which is satisfied in the vacuum.
The domain of $\theta$ can be analytically extended to $\theta \in (-\infty,\infty)$ by piecing together each of the branches of the Mexican hat, which have a change in potential energy of $\Delta V = g (2 \pi)^p$ for every winding.  This symmetry breaking mechanism has also been invoked as a means to dynamically lower the value of the cosmological constant~\cite{Abbott:1984qf} and is well motivated from effective field theory independently of the monodromy mechanism we study here.

With Eq.~\eqref{eqn:mono} large-field inflation can be realized by the canonically normalized, pseudo--Nambu-Goldstone boson $\phi=\theta \, f$ when $|\Phi|=f$, if
the initial angular displacement is sufficiently large $\theta_0 \gtrsim 10 \, \mpl/f$.
For $p=2$ this gives an effective inflaton mass of $m_\phi^2 = 2g /f^2$, which is less than the Hubble scale $H^2 \sim V_\mathrm{ASB}$ independently of $g$ and is sub-Planckian if $g \ll f^2$.  To match the amplitude of scalar perturbations as estimated from \emph{Planck}~\cite{Ade:2015xua}, the symmetry breaking parameter should be approximately $g \sim 10^{-9} f^p$, although the exact value depends on $\theta_0$ and $p$.

To ensure that the mass of the radial direction is $m_\rho \gtrsim H$ when the pivot scale leaves the horizon approximately $N_* \sim 55$ $e$-folds before the end of inflation we further require
\begin{equation}
  g \lesssim 4 f^2 \lambda \left( \frac{f}{\phi_0} \right)^p \sim f^{2+p},
  \label{eqn:superH}
\end{equation}
where we have removed $\mathcal O(1)$ terms in the second approximation.  A super-Hubble radial mass keeps the evolution close to the vacuum $\vev{\Phi} =f$ during inflation, which is an important requirement for this paper.\footnote{Ref.~\cite{Achucarro:2015rfa} has shown that the dynamics of the heavy radial direction can reduce the speed of sound of the inflaton fluctuations and suppress the amount of gravitational waves.  In order to isolate the predictions resulting from Planck-suppressed operators we do not include these effects in this paper.}

For a concrete example, Refs~\cite{Li:2014vpa,Li:2014unh,Li:2015taa} obtain monodromy in the phase of a complex scalar with a supergravity model where the global $U(1)$ symmetry of the K\"ahler potential is broken by the superpotential in such a way that it can give a variety of inflationary potentials $V(\phi)$, including Eq.~\eqref{eqn:mono}.  Refs~\cite{Berg:2009tg,McDonald:2014oza,McDonald:2014nqa,Barenboim:2014vea} have similar helical potentials with a range of phenomenological predictions.
Although realistic scenarios must account for analytical control over the symmetry breaking term at high energies, as well as the stability of a global minimum for reheating, we do not worry about these considerations here.

In addition to the Mexican hat shape of Eq~\eqref{eqn:mexhat} we can also add any operator that has the form
\begin{equation}
  V_m = c \left(\frac{\hat{\mathcal O}_m}{\Lambda^{m-4}}\right) ,
  \label{eqn:VV_planck}
\end{equation}
where $c$ is a dimensionless complex coupling, $\Lambda$ is the cutoff scale, and
\begin{equation}
  \hat{\mathcal O}_m \equiv | \Phi|^{2n} \left( \Phi^{*,q} + \Phi^q \right).
  \label{eqn:V_planck}
\end{equation}
The operator $\hat {\mathcal O}_m$ has energy dimension
\begin{equation}
m=2n+q
  \label{eqn:m}
\end{equation}
and breaks the translational symmetry $\theta \to \theta + \mathrm{const.}$ in Eq.~\eqref{eqn:mexhat}, leaving a residual discrete symmetry $\theta \to \theta + 2 \pi/ q$.
All operators of this type are generally expected from effective field theory considerations at energy dimension $m \ge 5$.
With the cutoff scale set to $\Lambda = \mpl$, we interpret these operators as gravitational corrections to the renormalizable theory at sub-Planckian energies.\footnote{While the operators in Eq.~\eqref{eqn:V_planck} are built with Fourier series, we could use an alternative basis, \emph{e.g.}, the parametrization $\Phi = \chi + i \psi$ has operators $V_m = c \chi^n \psi^q$ that are also periodic in $\theta$ with a similar phenomenology to Eq.~\eqref{eqn:VV_planck}.  However, we ignore those bases that have higher order terms that do not leave a residual angular symmetry, as well as any possible derivative terms.}

The impact of operators like Eq.~\eqref{eqn:V_planck} has been well studied in the context of the Peccei-Quinn axion~\cite{Peccei:1977hh}, where
$\theta$ is coupled to the QCD gluon field and setting the vacuum $\vev{\theta} \to 0$ dynamically solves the strong-CP problem.
Experimental constraints on CP violation require $\vev{\theta} \lesssim 10^{-10}$ and
non-renormalizable terms like Eq.~\eqref{eqn:V_planck} force $\vev{\theta} \ne 0$, giving a generally CP-violating vacuum in the absence of extreme fine-tuning in the relative phases of the operators~\cite{Holman:1992us,Barr:1992qq,Ghigna:1992iv,Kamionkowski:1992mf,Kallosh:1995hi}.

\begin{figure}
  \includegraphics[width=246pt]{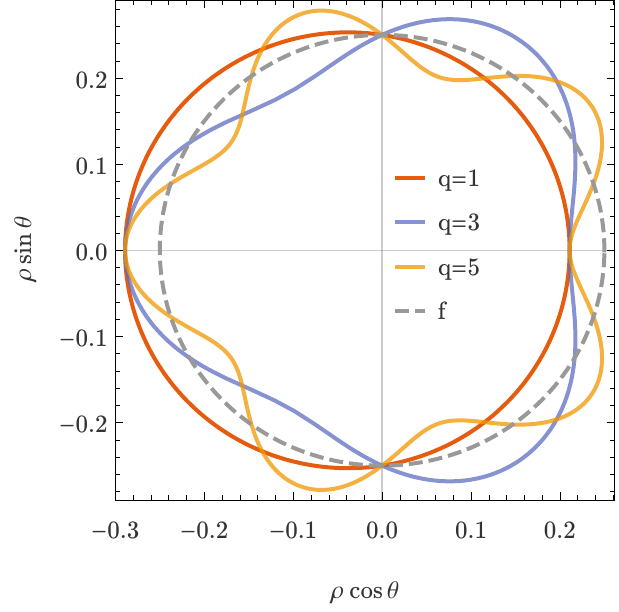}
  \caption{Value of $\vev{\Phi}$  when corrected by Eq.~\eqref{eqn:rhomin} in comparison to $f$.  The parameters are $\lambda =1$, $f=1/4$, and $|c|=1$.  We show the cases (\emph{red}) $q=1$, $n=2$; (\emph{blue}) $q=3$, $n=1$; and (\emph{yellow}) $q=5$, $n=0$.}
  \label{fig:vevplot}
\end{figure}

Typically, for inflation driven by the slowly rolling field $\theta$, an operator $V_{m\ge 5}$ is negligible since it is at most an $\mathcal O(f^5/\Lambda)$ correction to $V$ in the vacuum $\vev{\Phi}= \rho_\min = f$.
However, in the absence of Planck-suppressed operators like Eq.~\eqref{eqn:VV_planck}, the local minimum $\vev{\Phi}=\rho_\min$ is stabilized by the $|\Phi|^4$ term in the Mexican hat potential~\eqref{eqn:mexhat}.
Consequently, the operators $\hat{ \mathcal O}_5$ alter the minimum of $V$ in the radial direction of $\Phi$ at first order in $f$.
Operators of higher energy dimension $m >4$ change the location of the vacuum as
\begin{equation}
  \rho_\min \approx f \left[1 - \frac{m}{8} \left( \frac{|c|}{\lambda} \right) \cos \left( q \, \theta + \delta \right) \, \left(\frac{f}{\Lambda}\right)^{m-4} \right],
  \label{eqn:rhomin}
\end{equation}
where $\delta$ is the phase of the coupling $c$ and we have ignored terms $\mathcal O\left((\rho -f)^2\right)$ and terms higher order in $f$.
Fig.~\ref{fig:vevplot} shows $\rho_\mathrm{min}$ for different values of $n$ and $q$, with the decay constant $f$ set to a large value that has been exaggerated  in order to visualize the change in shape of the vev.

If we assume that $\rho = \rho_\min$, in addition to the negligible contributions from $V_{m \ge 5}$, the deviation of the vacuum geometry from $f$ in Eq.~\eqref{eqn:rhomin} affects the kinetic energy of the complex scalar as
\begin{equation}
  \partial_\nu \Phi^* \partial^\nu \Phi = \left[ \left( \frac{d \rho_\min}{d \theta} \right)^2 + \rho_\min^2 \right] \partial_\nu \theta \partial^\nu \theta.
  \label{eqn:XXX}
\end{equation}
We define a scalar field $\phi$ with a canonical kinetic energy in the usual fashion:
\begin{equation}
  \phi \equiv \int d \theta \, \sqrt{ \left( \frac{d \rho_\min}{d \theta} \right)^2 + \rho_\min^2},
  \label{eqn:phi_int}
\end{equation}
which is valid up to an arbitrary constant.

For simplicity we focus on a single leading order non-renormalizable operator with energy dimension $m=5$.  The derivative term in Eq.~\eqref{eqn:phi_int} for this operator is $\rho_\min' \sim \mathcal O(f^4)$ and can be effectively ignored compared to the contributions from $\rho_\min^2$.  Up to $\mathcal O(f^3)$ the integrand reduces to
\begin{equation}
  \phi \approx f \int d \theta \, \left[ 1 - \mu \, \frac{f}{\Lambda} \cos \left(q \, \theta+ \delta \right) \right] + \mathcal O(f^3),
  \label{eqn:XXX}
\end{equation}
where we have defined the dimensionless constant
\begin{equation}
  \mu \equiv \frac{5}{8} \left( \frac{|c|}{\lambda} \right),
  \label{eqn:mu}
\end{equation}
which we expect to be $\mathcal O(1)$ although it could be larger if $\lambda \ll 1$.
Assuming that the scale of $\mu$ does not disrupt the perturbative analysis with respect to $f$, the inflaton can be approximated as
\begin{equation}
  \phi \approx f \left[ \theta - \frac{\mu \, f}{q \, \Lambda} \sin \left( q \, \theta + \delta \right) \right] + \mathcal O \left(\frac{f^3}{\Lambda^2}\right),
  \label{eqn:phi_ptb_approx}
\end{equation}
which we can invert order-by-order in $f$ to get
\begin{equation}
  \theta \approx \frac{\phi}{f} + \frac{\mu \, f}{q \, \Lambda} \sin \left( q \, \frac{\phi}{f} + \delta \right) + \mathcal O \left(\frac{f^2}{\Lambda^2}\right).
  \label{eqn:theta_min}
\end{equation}
In the limit that $\mu \to 0$ we recover the standard definition $\theta = \phi/f$.

\begin{figure}
  \includegraphics[width=246pt]{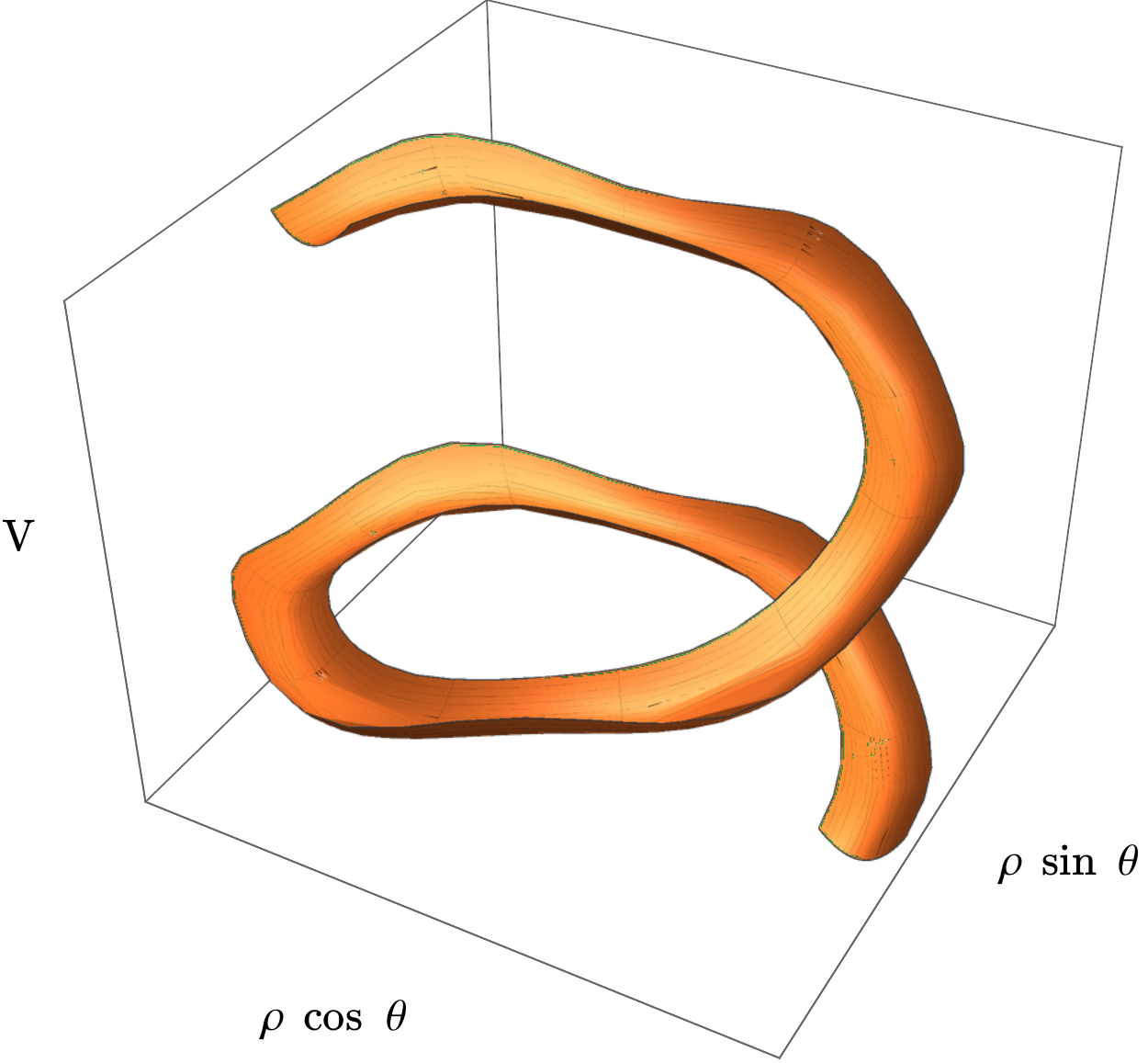}
  \caption{The portion corresponding to $0<\theta<3 \pi$ of the modified monodromy potential with one additional Planck-suppressed operator of energy dimension $m=5$, with $n=0$ and $q=5$.}
  \label{fig:3dvev}
\end{figure}

Consequently, in the low-energy description we can fix $\rho = \rho_\min$ and replace all instances of $\theta$ in the scalar sector of the action by the relationship given in Eq.~\eqref{eqn:theta_min}.  Defining the first and second terms in Eq.~\eqref{eqn:theta_min} as $\theta \equiv \theta_0 + \delta \theta_1$, the potential can be expanded like
\begin{equation}
  V(\theta) = V(\theta_0) + V'(\theta_0) \, \delta \theta_1.
  \label{eqn:XXX}
\end{equation}
For the combined Mexican hat action~\eqref{eqn:mexhat} and angular symmetry breaking potential~\eqref{eqn:mono}, the effective potential at tree level with the Planck-corrected inflaton in Eq.~\eqref{eqn:phi_ptb_approx} becomes
\begin{equation}
  V = g \left( \frac{\phi}{f} \right)^p + \frac{ g \, p \,\mu \, f}{q \, \Lambda} \left( \frac{\phi}{f} \right)^{p-1} \sin \left( q \, \frac{\phi}{f} + \delta \right)
  \label{eqn:mono_mod}
\end{equation}
plus terms higher order in $f$.  Fig.~\ref{fig:3dvev} shows this potential for $q=5$.
Since it is the most conservative choice we use a Planckian cutoff scale $\Lambda = \mpl =1$ for the rest of the paper.

The second term in Eq.~\eqref{eqn:mono_mod} does not come directly from the Planck-scale operators~\eqref{eqn:V_planck}, but rather from the substitution $\theta(\phi)$ into Eq.~\eqref{eqn:mono}, which is due to the modified shape of the vacuum in Eq.~\eqref{eqn:rhomin}.
The phase $\delta$ is important, since any of the possible oscillatory terms from Planck-suppressed operators and any oscillations coming from instanton effects generally have different relative phases, which we discuss in Sect.~\ref{ssect:mult_diff}.
With $p=1$ and $\mu \sim \mathcal O(1)$, the ratio $q/f$ fixes both the scale and frequency of the new oscillatory term compared to the angular symmetry breaking scale.

\section{A non-perturbative elliptical vacuum}
\label{sect:ellipse}

For sub-Planckian field values the corrections to the shape of the vacuum from non-renormalizable operators like Eq.~\eqref{eqn:V_planck} are naturally small.  However, we show in this section that the corrections to the potential $V$ can be kept perturbatively small even when the vev is altered non-perturbatively.

For a simple example with a non-circular vev we consider the case where $\vev{\Phi}$ is an ellipse with semi-major axis $f$ and eccentricity $\xi$, which are related by the semi-minor axis $b$ as
\begin{equation}
  \xi^2 = 1 - \left( \frac{b}{f} \right)^2.
  \label{eqn:XXX}
\end{equation}
Instead of the Mexican hat potential~\eqref{eqn:mexhat}, we assume that $V(\Phi)$ is structured so that $\beta \to \beta + \mathrm{const.}$ keeps the potential energy constant and setting $\xi \to 0$ restores the $\theta \to \theta +\mathrm{const.}$ symmetry.
We leave the exact form of this potential implicit since we will only worry about the dynamics when $\Phi = \vev{\Phi}$.
In analogy to the perturbative corrections to the vacuum in Sect.~\ref{sect:vac}, we keep a canonical kinetic energy term for $\Phi$, which breaks the $\beta$-translational symmetry for the Lagrangian when $\xi \ne 0$.

We parametrize the complex scalar $\Phi$ in terms of a radial coordinate $\rho$ and an angular coordinate $\beta$ as
\begin{equation}
  \Phi = A \cosh \left( \rho + i \beta \right).
  \label{eqn:XXX}
\end{equation}
Surfaces of constant $\rho$ are ellipses with foci at $\pm A$ and we define the vev so that it follows one of these ellipses,
\begin{equation}
  \vev{\Phi} = \rho_\min \equiv \tanh^{-1} \left( \frac{b}{f} \right).
  \label{eqn:XXX}
\end{equation}
We assume that the effective mass in a direction locally orthogonal to $\vev{\Phi}$ is sufficiently high so that we can treat the system as if it were one dimensional at low energies, with $\rho$ non-dynamical.
Near the vacuum we can therefore fix the location of the foci at $A = f \, \xi$.

\begin{figure}
  \includegraphics[width=246pt]{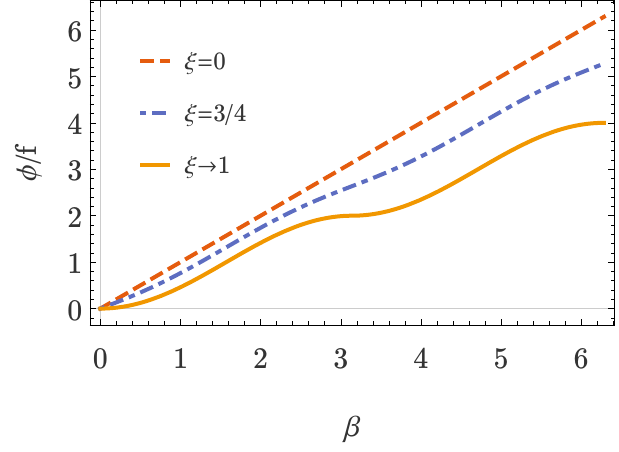}
\caption{The canonically normalized inflaton $\phi$ as a function of the elliptical angular variable $\beta$, where the vev has semi-major axis $f$ and eccentricity $\xi$.  The dependence $\phi(\beta)$ is invertible but non-elementary.}
  \label{fig:phi_of_beta}
\end{figure}

The canonical kinetic energy term for $\Phi$ in the vacuum becomes
\begin{equation}
  \partial_\nu \Phi^* \partial^\nu \Phi = \left( f^2 \sin^2 \beta + b^2 \cos^2 \beta \right) \partial_\nu \beta \partial^\nu \beta.
  \label{eqn:XXX}
\end{equation}
As in Eq.~\eqref{eqn:phi_int} we define the canonical inflaton as
\begin{equation}
  \phi = \int \sqrt{f^2 \sin^2 \beta + b^2 \cos^2 \beta} \; d \beta,
  \label{eqn:XXX}
\end{equation}
which can be expressed in terms of the incomplete elliptic integral of the second kind $\mathcal E$ as
\begin{equation}
  \phi = f \sqrt{1-\xi^2} \; \mathcal E \left[ \beta; \frac{\xi^2}{\xi^2-1} \right].
  \label{eqn:phi_of_beta}
\end{equation}
Fig.~\ref{fig:phi_of_beta} shows this relationship for a range of eccentricities $0\le \xi < 1$.  Although the function $\phi(\beta)$ is bijective, its inverse is not elementary due to the special function $\mathcal E$.  However, we can use the properties of $\mathcal E$ to obtain an approximation to $\beta(\phi)$ and calculate the potential energy of the canonical inflaton $\phi$.

For $\beta>0$, $\mathcal E$ is quasi-periodic with period $\pi$, satisfying
\begin{equation}
  \mathcal E \left[ n \pi; y\right] = 2n \; \mathcal E\left[  \pi; y\right],
  \label{eqn:EI_periodic}
\end{equation}
and $\mathcal E (0; y) = 0$ where $n$ is an integer and $y$ is an arbitrary parameter.
This relationship ensures that $\phi(\beta)$ is composed of a linear and an oscillatory component, which Fig.~\ref{fig:phi_of_beta} shows are approximately independent even as $\xi \to 1$.
Motivated by this, we decompose $\phi(\beta)$ in Eq.~\eqref{eqn:phi_of_beta} into a linear and oscillatory portion as
\begin{equation}
  \frac{\phi}{f} = \gamma_1 \; \beta + \mathrm{oscill.}
  \label{eqn:phi_lin_osc}
\end{equation}
where Eqs~\eqref{eqn:phi_of_beta}~and~\eqref{eqn:EI_periodic} give the pre-factor
\begin{equation}
  \gamma_1 \equiv \frac{1}{\pi} \sqrt{1 - \xi^2} \; \mathcal E \left[ \pi; \frac{\xi^2}{\xi^2-1} \right].
  \label{eqn:gamma1}
\end{equation}
This coefficient is bounded in the range $2/\pi < \gamma_1 \le 1$ for $0 \le \xi <1$.  To determine the form of the oscillatory term in Eq.~\eqref{eqn:phi_lin_osc} we examine the limiting behavior of the elliptic integral.

For small eccentricities $\xi \ll 1$ we can expand the elliptical integrals $\mathcal E$, which appear in $\gamma_1$ and in Eq.~\eqref{eqn:phi_of_beta}, to get
\begin{equation}
  \frac{\phi}{f} \approx \left( 1 - \frac{1}{4} \xi^2 \right) \beta -  \left(\frac{\xi^2}{8}\right) \sin 2 \beta + \mathcal O(\xi^4).
  \label{eqn:small_ecc}
\end{equation}
We can compare this directly to the perturbative case of Eq.~\eqref{eqn:phi_ptb_approx} and see that the functional form of these expressions matches in this limit, as expected.
Similarly, for large eccentricities $\xi \to 1$, Eq.~\eqref{eqn:phi_of_beta} becomes
\begin{equation}
  \frac{\phi}{f} \approx
  \frac{2 }{\pi } \beta
  -\left(\frac{\sqrt{2}-1}{2}\right) \sin 2 \beta
  + \mathcal O(\xi^2 -1).
  \label{eqn:large_ecc}
\end{equation}
Therefore, Eq.~\eqref{eqn:large_ecc} has an identical functional form for the oscillatory term as the small-eccentricity expression Eq.~\eqref{eqn:small_ecc} at first order in $\xi^2$.  Since Fig.~\ref{fig:phi_of_beta} shows that intermediate eccentricities have similar behavior, a close approximation to Eq.~\eqref{eqn:phi_of_beta} for all eccentricities $0 \le \xi <1$ is
\begin{equation}
  \frac{\phi}{f} \approx \gamma_1 \; \beta + \gamma_2 \; \sin 2 \beta,
  \label{eqn:phiapprox}
\end{equation}
where we have defined the coefficient
\begin{equation}
  \gamma_2 \equiv \sqrt{1-\xi^2} \, \Delta \mathcal E
  \label{eqn:XXX}
\end{equation}
in terms of the change in the elliptical integral, which is
\begin{equation}
  \Delta \mathcal E \equiv \mathcal E \left[\frac{\pi}{4}; \frac{\xi^2}{\xi^2-1} \right] - \frac{1}{2} \mathcal E \left[\frac{\pi}{2}; \frac{\xi^2}{\xi^2-1} \right].
  \label{eqn:XXX}
\end{equation}
The coefficient $\gamma_2$ for the oscillatory term is bounded in the range $0 \ge \gamma_2 \gtrsim -0.21$ for eccentricities $0 \le \xi <1$.

Eq.~\eqref{eqn:phiapprox} is valid for any ellipse and is not restricted to a perturbative expansion in $\xi$.  Although this expression could be inverted numerically to obtain $\beta(\phi)$, the inverse expression $\beta(\phi)$ is transcendental.  An analytical inversion, as in Eq.~\eqref{eqn:theta_min}, can only be made once a suitably small parameter is identified.  Rather than restricting our analysis to small eccentricities, which would replicate the results of Sect.~\ref{sect:vac}, we instead solve for $\beta(\phi)$ perturbatively in the parameter
\begin{equation}
  \gamma (\xi) \equiv - \frac{\gamma_2}{\gamma_1}.
  \label{eqn:XXX}
\end{equation}
This ratio is naturally small since the oscillatory term in Eq.~\eqref{eqn:phiapprox} is sub-dominant to the linear even in the large-eccentricity limit~\eqref{eqn:large_ecc}.  Importantly, $\gamma$ satisfies the exact bounds
\begin{equation}
  0 \le \gamma \le \frac{\pi}{4} \left(\sqrt{2}-1 \right) \lesssim \frac{1}{3},
  \label{eqn:XXX}
\end{equation}

\begin{figure}
  \includegraphics[width=246pt]{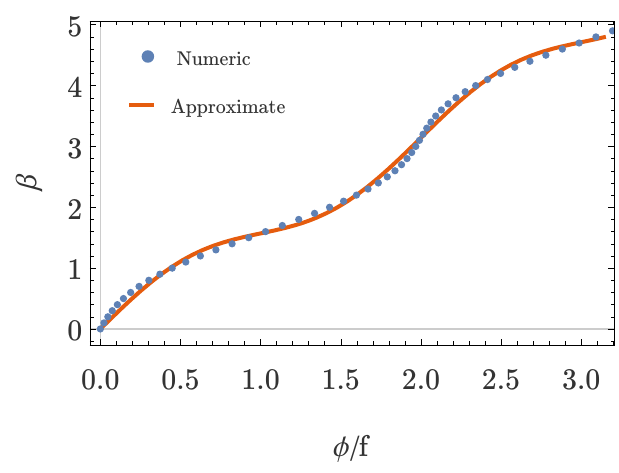}
  \caption{The inflaton $\phi$ in terms of the angular variable $\beta$ when the expression \eqref{eqn:phiapprox} is inverted both numerically and approximately with Eq.~\eqref{eqn:betaapprox}.  This shows the worst-case scenario with maximum eccentricity $\xi \to 1$, with good agreement between the numeric and approximate solutions.}
  \label{fig:beta}
\end{figure}

The separation in scales between the linear and oscillatory contributions to $\phi$ justifies an order-by-order perturbative solution to the inverse function $\beta(\phi)$.  At first order this yields
\begin{equation}
  \beta \approx \frac{\phi}{\bar f } + \gamma \; \sin \left( \frac{2 \, \phi}{\bar f} \right),
  \label{eqn:betaapprox}
\end{equation}
where the effective decay constant is
\begin{equation}
  \bar f (\xi) \equiv f \; \gamma_1,
  \label{eqn:XXX}
\end{equation}
which is fixed for a given eccentricity.  Eq.~\eqref{eqn:betaapprox} is analogous to Eq.~\eqref{eqn:theta_min} in Sect.~\ref{sect:vac}.
Fig.~\ref{fig:beta} compares this perturbative inverse function to the exact numerical inverse in the maximum eccentricity case $\xi \to 1$.  In this limit the ratio $ \gamma = - \gamma_2/\gamma_1$ maximizes and the approximation is close to the numerical values even for this most extreme case.  Fig.~\ref{fig:beta} only shows $\beta \in [0,\pi]$, but the periodicity of the elliptical integral in Eq.~\eqref{eqn:EI_periodic} ensures this approximation is valid for all values of $\beta$.

Given the relationship~\eqref{eqn:betaapprox} we can replicate the analysis of Sect.~\ref{sect:vac}, substituting $\beta(\phi)$ into the low-energy effective potential.
In particular, since we constructed this potential so that it has the symmetry $\beta \to \beta + \mathrm{const.}$, we assume that the angular symmetry breaking mechanism acts as
\begin{equation}
  V_\mathrm{ASB} = g \; \beta^p
  \label{eqn:XXX}
\end{equation}
rather than the expression in Eq.~\eqref{eqn:mono}.
Therefore, the elliptical vev yields a correction to the potential as
\begin{equation}
  V(\phi) \approx g \left( \frac{\phi}{\bar f} \right)^p \left[1 + p \, \gamma \; \left( \frac{\bar f}{\phi} \right) \sin \left(\frac{2\phi}{\bar f} \right) \right]
  \label{eqn:V_ellipse}
\end{equation}
to first order in $\gamma$.  This expression closely matches Eq.~\eqref{eqn:mono_mod}, which arises from modifying the geometry of the vev $\vev{\Phi} = \rho_\min(\theta)$ from adding non-renormalizable operators to the Lagrangian.  However, the frequency of the oscillation term in Eq.~\eqref{eqn:V_ellipse} is not possible with an operator of energy dimension $m=5$.  Therefore the oscillatory term above is not Planck-suppressed.  Instead, large elliptical deviations from a circular vev should come from relevant or marginal operator.  We leave the exact construction of this type of scenario ambiguous.

This toy model demonstrates that even non-perturbatively modified vacuum geometries can still yield easily categorizable alterations to the basic shape of the potential.  As long as we can identify an angular parameter $\{\theta, \beta, \dots \}$ that  varies over some finite range in the absence of monodromy, then the relationship $\phi(\theta)$ in the modified vacuum should also be periodic and invertible, since Fourier series like Eq.~\eqref{eqn:V_planck} are the natural basis with which to build operators over a compact subset of the real numbers.  However, one might expect that being able to identify perturbative parameters that give a tractable perturbative inverse function $\theta(\phi)$ is model dependent.

\section{Power spectrum features from non-circular monodromy}
\label{sect:pk}

Changing the shape of the vev in Sects~\ref{sect:vac}~and~\ref{sect:ellipse} has resulted in a modified monodromy potential of the form
\begin{equation}
  V(\phi) = g \left( \frac{\phi}{f} \right)^p \left[1 + p\; \kappa \; \left( \frac{f}{\phi} \right) \sin \left(\frac{q \; \phi}{ f} \right) \right],
  \label{eqn:V_phi}
\end{equation}
where the constants $f$, $q$, and $\kappa$ are related to the parameters in the potentials of Eqs.~\eqref{eqn:mono_mod}~and~\eqref{eqn:V_ellipse} in the obvious way.
The ratio $g/f^p$ sets the amplitude of scalar perturbations and $\kappa$ controls the oscillatory contribution to the primordial spectra.
We estimate the power spectrum resulting from Eq.~\eqref{eqn:V_phi} using the $\delta N$ formalism.  This relates the curvature perturbation $\zeta$ on equal density hypersurfaces to the field perturbations $\delta \phi$ on spatially flat hypersurfaces via
\begin{equation}
  \zeta = N' \; \delta \phi
  \label{eqn:dN}
\end{equation}
to first order in $\delta \phi$, where $N$ is the number of $e$-folds between the flat and equal density hypersurfaces and $N' = dN/d\phi$.

We approximate the power spectrum whenever a mode $k$ crosses the horizon by
ignoring any non-standard sub-horizon evolution for the Fourier transformed field values $\delta \phi_k$.  This gives the standard result for the power spectrum of scalar perturbations
\begin{equation}
  \mathcal P_{\delta \phi}(k) = \frac{V_k}{12 \pi^2},
  \label{eqn:XXX}
\end{equation}
where $V_k = V(\phi_k)$.  From Eq.~\eqref{eqn:dN} the primordial curvature power spectrum is
\begin{equation}
\mathcal P_\zeta(k) = \left( N' \right)^2 \mathcal P_{\delta \phi}(k).
  \label{eqn:Pk}
\end{equation}
Using the slow-roll approximation, the number of $e$-folds of expansion can be calculated as
\begin{equation}
  N = - \int_{\phi_k}^{\phi_c} \frac{V}{V'} d\phi,
  \label{eqn:N_sr}
\end{equation}
where $\phi_c$ is the value of the field at the end of inflation and $\phi_k$ is the field value when $k=aH$.
The derivative of Eq.~\eqref{eqn:N_sr} is
\begin{align}
  \label{eqn:Nprime_k}
  N' = - \frac{\phi_k}{p} \left[1 - q \,  \kappa \cos \left( \frac{ q \, \phi_k}{ f} \right)  \right. \\
  \left. + q \,  \kappa \left( \frac{  f}{\phi_k}\right) \sin \left( \frac{q \,  \phi_k}{ f}\right) \right], \notag
\end{align}
where we have disregarded terms that are $\mathcal O(\kappa^2)$.  The third term in the square brackets can be ignored for $\phi_k \gg f$, which is the case for those modes that leave the horizon within a few $e$-folds of the pivot scale $k_*=0.002 \, \impc$, which has $\phi_* \sim 10$ for $p \sim 1$.\footnote{The derivatives $\partial \phi_c/\partial \phi_k$ vanish for adiabatically evolving scalar fields.}

We estimate the horizon crossing value $\phi_k$ of the inflaton field  in Eq.~\eqref{eqn:Nprime_k} using only the non-oscillatory portion of Eq.~\eqref{eqn:V_phi}, \emph{i.e.}, with $\kappa \to 0$.  This is only a rough approximation to the slow-roll dynamics for $\phi$ and results in a value for $N'$  in Eq.~\eqref{eqn:Nprime_k} that artificially suppresses the oscillatory terms, since we are ignoring a component to $\phi_k$ that is in phase with these terms.
A more accurate estimate of $\phi_k$ might be obtained by replicating the techniques\footnote{This becomes more difficult when the angular symmetry breaking term does not have the exponent $p=1$.} of Refs~\cite{Flauger:2009ab,Flauger:2014ana} or using generalized slow-roll techniques~\cite{Stewart:2001cd,Gong:2002cx,Lee:2005bb}.
However, this approximation suffices to understand the general behavior of the power spectrum as a function of scale.

Setting $\kappa = 0$ reduces Eq.~\eqref{eqn:N_sr} to
\begin{equation}
  N \approx - \frac{1}{2p} \phi_k^2  + \mathcal O\left( \frac{\phi_c^2}{\phi_k^2} \right).
  \label{eqn:N_vbeta}
\end{equation}
Since the power law term in Eq.~\eqref{eqn:V_phi} gives large-field inflation, we ignore any contributions to $N$ that result from $\phi_c$ throughout the rest of this section.
For the pivot scale, Eq.~\eqref{eqn:N_vbeta} gives
\begin{equation}
\phi_* \approx \sqrt{2p N_*}.
  \label{eqn:XXX}
\end{equation}

We relate $\phi_k$ to the field value $\phi_*$ when the pivot scale  leaves the horizon $N_* \approx 55$ $e$-folds before the end of inflation by
\begin{equation}
  \frac{k}{k_*} = \frac{a_k H_k}{a_* H_*} = e^{N_* - N} \left( \frac{\phi_k}{\phi_*} \right)^{p/2},
  \label{eqn:k_relation}
\end{equation}
where we have used the slow-roll approximation $H_k^2 \sim V_k$ and have related the scale factors at horizon crossing to the scale factor at the end of inflation by $a_k =a_c \, e^{-N}$.
Using Eq.~\eqref{eqn:N_vbeta}, this can be expressed as
\begin{equation}
  N_* \left( \frac{\phi_k}{\phi_*} \right)^2 + \frac{p}{2} \, \log \frac{\phi_*}{\phi_k} = N_* - \log \frac{k}{k_*}.
  \label{eqn:N2}
\end{equation}
Modes that leave the horizon within a few $e$-folds of each other have only a small change in $\phi_k$.  Consequently, we solve this equation for $\phi_k$ while ignoring $\log \phi_*/\phi_k$ in order to understand the behavior of $P_\zeta(k)$ only near $k_*$.  The solution to Eq.~\eqref{eqn:N2} in this limit is
\begin{equation}
  \phi_k^2 \approx \phi_*^2 \left(1- \frac{1}{N_*} \log \frac{k}{k_*} \right).
  \label{eqn:phi_k}
\end{equation}
This expression is valid only for $k / k_* \ll e^{N_*}$ and when $\phi_k$ depends only weakly on the oscillatory terms in $V$.

\begin{figure}
  \includegraphics[width=246pt]{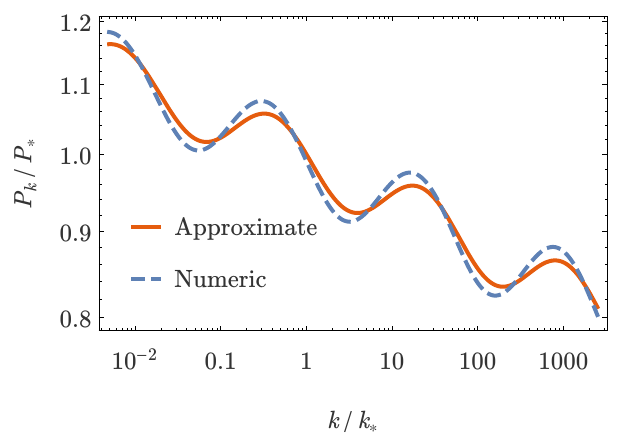}
  \caption{Primordial power spectrum $\mathcal P_\zeta(k)$ from the potential~\eqref{eqn:V_phi}, normalized by the amplitude of the power spectrum at the pivot scale $k_*$.  The parameters are $f=10^{-1}$, $\kappa=10^{-2}$, $N_*=55$, $p=2/3$, and $q=2$, which describes an elliptical vacuum of eccentricity $\xi = 0.28$.  The solid orange line is the $\delta N$ approximation and the dashed blue line is the exact numerical value.}
  \label{fig:pk}
\end{figure}

Substituting Eqs~\eqref{eqn:Nprime_k}~and~\eqref{eqn:phi_k} into Eq.~\eqref{eqn:Pk} gives an estimate of the primordial power spectrum.
Fig.~\ref{fig:pk} compares this approximation to an exact numerical calculation for the potential~\eqref{eqn:V_phi} using the publicly available inflation solver \textsc{MultiModeCode}~\cite{Mortonson:2010er,Easther:2011yq,Norena:2012rs,Price:2014xpa}.  \textsc{MultiModeCode} numerically solves the Klein-Gordon and mode equations for $\phi$ on sub-horizon scales and is exact to first order in $\delta \phi_k$.  The chosen parameters for Fig.~\ref{fig:pk} correspond to an elliptical vev with eccentricity $\xi =0.28$, which results in sinusoidal oscillations in $\mathcal P_\zeta$ as a function of $\log k/k_*$.  The analytic solution is a close approximation, although the amplitude of the oscillations is suppressed relative to the exact calculation, as expected.  For smaller amplitude oscillations with $|\kappa| \lesssim 10^{-2}$ the approximation becomes a better fit.

Constraints on primordial tensor modes from \emph{Planck} and BICEP2~\cite{Ade:2013rta,Ade:2013ydc,Ade:2015lrj,Ade:2015tva} restrict $p \lesssim 2$, although this is model dependent.
To match recent studies that constrain oscillations in the primordial spectrum~\cite{Peiris:2013opa,Easther:2013we,Meerburg:2013cla,Meerburg:2013dla}, we generally expect $|\kappa| \lesssim 10^{-3}$ and $f \lesssim 10^{-2}$ for the fiducial case of $p=1$.
When Eq.~\eqref{eqn:V_phi} is generated by Planck-suppressed operators, the upper limit on $|\kappa|$ corresponds to $\mu \lesssim 0.5$ with $\mu$ defined as in Eq.~\eqref{eqn:mu}.\footnote{The data can allow much wider ranges for the parameters, since the posterior probabilities for the model parameters can be highly non-trivial.  Furthermore, large wavelength oscillations in $\mathcal P_\zeta(k)$ can be confused for models that have large (positive or negative) running~\cite{Kobayashi:2010pz}, due to the finite range of multipole moments that can be measured at high accuracy in the cosmic microwave background.}

\section{Differentiating Planck-corrected and gauged oscillations}
\label{sect:diff}

\subsection{The toy case}
\label{ssect:toy_diff}

In Sects~\ref{sect:vac}~and~\ref{sect:ellipse} we considered an effective description of monodromy inflation where a global angular symmetry $\theta \to \theta + \mathrm{const.}$ is explicitly broken and irrelevant, Planck-suppressed operators.  This differs from many axion monodromy scenarios that come from explicit high energy considerations, such as string axions~\cite{Baumann:2014nda} or extranatural inflation~\cite{ArkaniHamed:2003wu}, where the translational symmetry in $\theta$ is a gauge symmetry.

Typically, the field $\theta$ in high energy theories is identified with an axion, which is obtained by integrating gauge fields over cycles in compact extra dimensions.  The residual gauge symmetry for this scalar severely restricts the allowed structure of the Lagrangian and precludes operators like those in Eq.~\eqref{eqn:V_planck} from perturbatively altering the shape of the vacuum $\vev{\Phi}$ once the gauge is fixed.
However, the gauge symmetry in these cases requires that $\Phi$ is coupled to gauge fields, which induces an oscillatory term to the potential via non-perturbative instanton effects, which take the form
\begin{equation}
  V_\mathrm{NP} = g_\mathrm{NP} \cos \left( \frac{\phi}{f} + \delta_\mathrm{NP} \right).
  \label{eqn:V_inst}
\end{equation}
We treat this as the prototypical potential for gauged monodromy inflation, although the model might be different in more specific cases, particularly those that incorporate a natural variability in the oscillation frequency~\cite{Berg:2009tg,McDonald:2014oza,Barenboim:2014vea,Flauger:2014ana}.

In principle, signals in the primordial spectrum due to sinusoidal terms like Eq.~\eqref{eqn:V_phi} should be distinguishable from the features derived exclusively from non-perturbative instanton effects like in Eq.~\eqref{eqn:V_inst}.
The difference is most obvious for the case where the vacuum geometry is modified by a single Planck-suppressed term in the Lagrangian and the potential is closely approximated by Eq.~\eqref{eqn:mono_mod}.
Most obviously, the coupling constant for the sinusoidal term varies as a function of $\phi$ for $p \ne 1$.  For $p<1$ modes $k$ that leave the horizon closer to the end of inflation feel a relatively larger effect from the oscillatory portion of Eq.~\eqref{eqn:mono_mod} than those that leave the horizon earlier.\footnote{For $\phi \ll f$ and $p<1$, the sinusoidal term dominates over the diverging pre-factor $(\phi/f)^{p-1}$ so that $V \to 0$ here.}
The resulting signal in the power spectrum for these smaller modes may be detectable in the correlation functions of large scale structure in large-volume surveys~\cite{Schlegel:2009uw,Abell:2009aa,Laureijs:2011gra,Chluba:2015bqa}, which are able to constrain the amplitude of small modes $k \gg k_*$.

The non-renormalizable oscillations also have a larger frequency than the instanton terms by an integer multiple,
\begin{equation}
  \omega_q \equiv \frac{q}{f} \qquad \mathrm{vs} \qquad \omega_\mathrm{NP} = \frac{1}{f},
  \label{eqn:freq}
\end{equation}
where $q \in \{1,3,5\}$ for the leading Planck-scale correction.  This gives features in $P_\zeta(k)$ that are of higher frequency than features from Eq.~\eqref{eqn:V_inst} for fixed values of $f$.

Furthermore, if we assume that all dimensionless couplings in Eq.~\eqref{eqn:mono_mod} are exactly unity, then the coupling for the new sinusoidal term from $V_5$ as compared to the leading order polynomial term is
\begin{equation}
  \frac{p \; \mu \; f}{q} \to \left(\frac{5}{8}\right) \frac{p}{ \omega_q}.
  \label{eqn:XXX}
\end{equation}
With the angular symmetry breaking exponent $p$ fixed, this only has the frequency $\omega_q$ as a free parameter.  The relative amplitude of oscillatory features in the primordial spectrum $P_\zeta(k)$ that result from the instanton and non-circular vev effects is then directly tied to their relative frequencies.  However, the amplitude of power spectrum oscillations that result from Eq.~\eqref{eqn:V_phi} is independent of those coming from Eq.~\eqref{eqn:V_inst}.

\subsection{Power spectrum with multiple oscillatory components}
\label{ssect:mult_diff}

In general, the analysis in Sect.~\ref{ssect:toy_diff} is overly simplistic, since we should allow for all possible non-renormalizable operators to contribute to the modification of the vacuum, including derivative terms that we have ignored throughout this paper.  However, adding only those Planck-suppressed operators of energy dimension $m=5$ that are described in Eq.~\eqref{eqn:V_planck} changes the relationship $\theta(\phi)$ in Eq.~\eqref{eqn:theta_min} to
\begin{equation}
  \theta = \frac{\phi}{f} + f \sum_{i=1}^3 \frac{\mu_i}{q_i} \sin \left(q_i \, \frac{\phi}{f} + \delta_i \right),
  \label{eqn:theta_mult}
\end{equation}
where the sum is over the three values of $q_i \in \{1,3,5 \}$ that are allowed from Eq.~\eqref{eqn:m}.  The coefficients $\mu_i$ are defined as in Eq.~\eqref{eqn:mu} for each of the $q_i$, where the phases $\delta_i$ are assumed to be uniformly distributed.  Eq.~\eqref{eqn:theta_mult} is a perturbative inverse function that is valid to $\mathcal O(f^2)$ and requires that the sum of the individual couplings is sub-dominant to the linear term $\theta=\phi/f$.  This marginally lowers the range of decay constants for which this expression is valid in comparison to Eq.~\eqref{eqn:theta_min}.
Potentials with multiple oscillatory terms have been studied in the absence of monodromy in Refs~\cite{Czerny:2014wza,Czerny:2014xja,Czerny:2014wua,Czerny:2014qqa} and with monodromy in Refs~\cite{Kobayashi:2014ooa,Higaki:2014sja}.  However, the tight relationships between the coupling constants and frequencies in Eq.~\eqref{eqn:theta_mult}, combined with the limited number of Planck-suppressed operators of energy dimension $m=5$, severely reduce the effective dimensionality of parameter space for this scenario in comparison to previous work.

\begin{figure}
  \includegraphics[width=246pt]{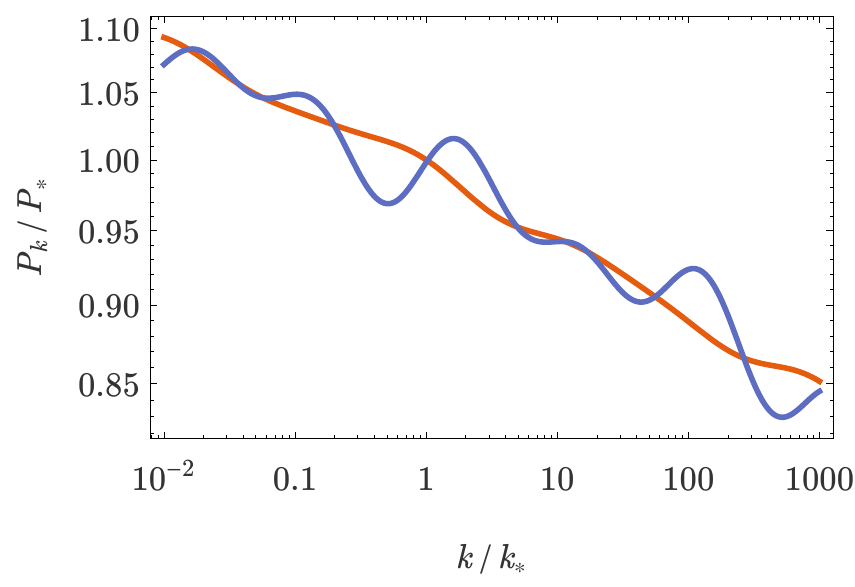}
  \caption{Power spectrum when all lowest-order Planck-suppressed operators of $\mathcal O(f^5)$ are included.  The angular symmetry breaking exponent is $p=1/3$ and the decay constant is $f=0.1$.  The relative phases between the different oscillatory contributions and their overall couplings are chosen independently from the distributions $\delta_i \sim \mathcal U[0,2\pi]$ and $\mu_i \sim \mathcal U[0,1]$. Two different selections of $\mu_i$ and $\delta_i$ are plotted.}
  \label{fig:pk_mult}
\end{figure}

The general expression~\eqref{eqn:theta_mult} modifies the potential~\eqref{eqn:V_phi} so that it contains a summation of out-of-phase oscillation terms.  This similarly affects the derivative of the number of $e$-folds as
\begin{equation}
  N ' \approx  -\frac{\phi_k}{p}  \left[1 - f^2 \sum_i \mu_i \cos \left( q_i \, \frac{ \phi_k}{f} + \delta_i \right) \right].
  \label{eqn:Nprime2}
\end{equation}
We again assume that the total contribution of the oscillatory terms above only negligibly affects the relationship $\phi_k(\phi_*)$ in Eq.~\eqref{eqn:phi_k}.  Therefore, the power spectrum $P_\zeta(k)$ can be estimated by simply substituting $N'$ from Eq.~\eqref{eqn:Nprime2} into Eq.~\eqref{eqn:Nprime_k}.
The power spectrum then roughly corresponds to a power-law with a superposition of sinusoidal terms, each of which has a form as in Fig.~\ref{fig:pk}.  Fig.~\ref{fig:pk_mult} displays the power spectrum for the superimposed signal with two different choices of couplings $\mu_i$ and phases $\delta_i$.
This relatively complex primordial signal differs substantially from the easily distinguishable oscillations expected from the potential~\eqref{eqn:V_inst} alone.

\section{Summary}
\label{sect:concl}

In Sect.~\ref{sect:vac} we studied the low-energy effective description of monodromy inflation resulting from the evolution of the angular component $\theta$ of a complex scalar field $\Phi$.  We started with the only renormalizable potential that gives a non-zero vev $\vev{\Phi} = f$ and assume $f \ll \mpl$.
Motivated by effective field theory we included higher order terms with energy dimension $m \ge 5$.  Although these terms are suppressed at order $\mathcal O(f^5)$, we showed that they modify the geometry of the vev so that it has an angular dependence that is affected more severely: $\vev{\Phi} \to f [1 + \mathcal O(f(\theta))]$.

When we restrict $\Phi = \vev{\Phi}$, the new geometry for the vacuum adds an angular dependence to the canonical kinetic energy term for $\Phi$.
In the absence of irrelevant operators, the canonically normalized scalar field that drives slow-roll inflation is $\phi = f \, \theta$, but acquires a sinusoidal dependence when these terms are included.  This correction is first order in $f$ and the inverse function is given schematically by $\theta \sim \phi/f + \mathcal O(f) \sin(\omega \phi)$.

In Sect.~\ref{sect:ellipse} we looked at more radical modifications to the shape of the vacuum.  We parametrized the vev as an ellipse with arbitrary eccentricity $0 \le \xi < 1$ and found that the elliptical angular coordinate variable $\beta$ is related perturbatively to the inflaton $\phi$ in terms of the small parameter $0 \le \gamma(\xi) < 1/3$, which is true for all possible eccentricities.  This parameter describes the ratio of linear and oscillatory contributions to the elliptic integral $\mathcal E (\pi; y)$, which is guaranteed to be small since $y(\xi) <0$ for any ellipse.  This gives a description of the effective potential in the vacuum that is analogous to Sect.~\ref{sect:vac}, without requiring a perturbatively small modification to the circular Mexican hat vev.

Since the relationship between the angular parameter $\theta$ and the inflaton $\phi$ has a first order correction, a term $V \sim \theta^p$ that breaks the symmetry $\theta \to \theta + 2 \pi$ is also corrected at first order in $f$.  The modified potential obtains one or more sinusoidal terms, which each contribute oscillations to the primordial power spectrum $P_\zeta(k)$.
In Sect.~\ref{sect:pk} we calculated the power spectrum resulting from these types of corrections and found that they are oscillations in $\log k$.

Without a reason to exclude Planck-suppressed operators from the theory, the existence of multiple oscillatory terms in the potential is completely general.
As discussed in Sect.~\ref{sect:diff}, the $P_\zeta(k)$ from many such terms should be closely approximated by the linear superposition of signals with varying phases.
However, the individual signals have a close relationship between their frequencies,
which is constrained to a few possible integer values due to the energy dimension of the non-renormalizable operators.  Furthermore, the overall coupling constant for the terms that determine the size of oscillations in $P_\zeta(k)$ can be related to the oscillation frequency, if the dimensionless parameters in the theory are $\mathcal O(1)$.

The EFT-motivated description of the low-energy theory of a complex scalar contrasts with those cases where monodromy inflation results from the simplest non-perturbative breaking of a gauge symmetry.  In principle, the effects of the modified vacuum mechanism presented here would distinguish a high energy axion monodromy theory from monodromy arising in an effective low-energy description.
Consequently, searches for non-sinusoidal features in the primordial power spectrum might eventually allow us to extract a Planckian signal from the primordial spectrum.  Conversely, the absence of complicated oscillations in $P_\zeta(k)$ would constrain the allowed form of Planck-suppressed operators in monodromy inflation, adding to existing constraints on high dimension operators from local non-Gaussianity~\cite{Assassi:2013gxa}.  However, obtaining significant evidence in favor of a non-sinusoidal feature in the primordial spectrum remains a challenging problem.

\acknowledgments

I thank Richard Easther, Dan Green, Hiranya Peiris, Eva Silverstein, Ewan Stewart, and Jonathan White for helpful discussions.
I acknowledge the use of the
New Zealand eScience Infrastructure (NeSI) high-performance computing facilities, which are funded jointly by NeSI's collaborator institutions and through the Ministry of Business, Innovation \& Employment's Research Infrastructure programme [{\url{http://www.nesi.org.nz}}]. This work has been facilitated by the Royal Society under their International Exchanges Scheme.

\bibliographystyle{JHEP}
\bibliography{newreferences,references}

\end{document}